\documentclass[singlecolumn,pras,amssymb]{revtex4-2}

\usepackage{natbib}
\usepackage{graphicx}
\usepackage{bm}
\usepackage{color}

\usepackage{float}
\usepackage{amsmath}
\usepackage{amsfonts}
\usepackage{amssymb}
\usepackage{makeidx}
\usepackage{graphicx}
\usepackage{placeins} 
\usepackage{color}
\usepackage{physics}
\usepackage{ragged2e}
\usepackage{booktabs}
\usepackage{multirow}
\usepackage{array}

\newcolumntype{C}[1]{>{\centering\arraybackslash}m{#1}} 
\newcolumntype{L}[1]{>{\raggedright\arraybackslash}m{#1}}
\usepackage{multirow}

\begin{document}	
	
		\title{Polaron effects on the information backflow  in Jaynes-Cummings model}
	\author{ Saima Bashir$^1$, Mehboob Rashid$^2$,  Rayees A Malla$^3$, Muzaffar Qadir Lone$^2$\footnote{corresponding author: lone.muzaffar@uok.edu.in} }	
\affiliation{ $^1$ Department of Physics, National Institute of Technology Srinagar, India 190006  \\$^2$Chennai Mathematical Institute, Siruseri Chennai, India 603103 \\
	$^3$ Quantum Dynamics Lab, Department of Physics, University of Kashmir Srinagar, India 190006 }
	
	\begin{abstract}
		We investigate the influence of  phonon degrees of freedom on the qubit dynamics in Jaynes-Cummings (JC) model. A strong qubit-phonon coupling is considered giving rise to Jaynes-Cummings-Holstein (JCH) model. Under anti-adiabatic conditions, we perform a unitary transformation to make the underlying problem tractable through Redfield-type non-Markovian master equation.  Analytical expression for the time-dependent coherence is obtained, incorporating both cavity-induced dissipation and phonon-induced dressing effects. The dynamics of JC model is highly non-Markovian for a narrow spectral width and finite detuning. However, a non-zero phonon coupling suppresses these non-Markovian features by effectively reducing the qubit-cavity interaction strength. {It is observed that polaronic dressing effectively supresses the detuning effects. Furthermore, the coherence-based non-Markovianity measure shows an order-of-magnitude suppression in the JCH model, indicating a new dynamical regime, while memory effects extend over a wider range of spectral densities than in the JC model.}

	\end{abstract}

	\maketitle
\section{Introduction}
  {Open quantum systems constitute a foundational framework in contemporary quantum science, as any realistic quantum device is inevitably coupled to its surrounding environment. Such system–environment interactions typically give rise to decoherence and dissipation, thereby imposing fundamental limitations on the performance and scalability of quantum technologies\cite{breuer2002theory,leggett1987dynamics,nielsen2010quantum}. However, when the environment exhibits internal structure or finite correlation times, the resulting dynamics may depart significantly from the standard Markovian approximation. In these regimes, a partial backflow of information from the environment to the system can occur due to memory effects, resulting in non-Markovian dynamics. The characterization, control, and exploitation of non-Markovian dynamics have therefore become central objectives across quantum optics, quantum information processing, and condensed-matter–inspired quantum platforms\cite{Haikka,Breuer2016}. 
   	
   	Within this broader context, light–matter interactions in photonic cavities provide a highly controllable and conceptually transparent setting for investigating open quantum dynamics with applications in quantum optics, quantum information, and molecular photonics} \cite{zubairy,vahala2003,nielsen2010quantum,Preskill}. In particular, systems where qubits are coupled to confined electromagnetic modes, as described by the Jaynes–Cummings (JC) and Tavis–Cummings (TC) models, have provided important insights into cavity quantum electrodynamics (QED) \cite{jj, shore1993jaynes,hou2024,tavis, garraway1997nonperturbative}. Beyond this, it has become increasingly evident that vibrational degrees of freedom play a crucial role in such systems, especially in solid-state qubits, superconducting circuits, and organic polaritonic devices. In these systems,  interactions with localized phonon modes can significantly alter energy transfer and decoherence dynamics\cite{nazir2016modelling, ebbesen2016hybrid, herrera2020molecular,Weiss2012}.	
	To capture these effects, hybrid models combining cavity and vibrational couplings have been introduced. The Holstein–Tavis–Cummings (HTC) model, for example, incorporates local vibrational modes coupled to molecular ensembles interacting with a cavity mode \cite{cwik2014polariton, herrera2016cavity, keeling2020bose}. Such models have revealed that vibrational degrees of freedom  not only induce decoherence but actively shape polaritonic transport and relaxation by creating phonon side-bands and mediating energy redistribution \cite{herrera2017absorption, luker2017phonon, chin2010exact}. These findings highlight the importance of treating vibrational modes as integral dynamical elements rather than passive decohering environments. Based on these developments, we study in this paper, Jaynes-Cummings model with a single mode  Holstein term. This configuration is often called as  Jaynes–Cummings–Holstein (JCH) model \cite{shore1993jaynes,larson2021jaynes,cai2021robust,hou2024}, which serves as a minimal framework for exploring how vibrational coupling modifies qubit-cavity dynamics in hybrid quantum systems. Such settings are particularly relevant for single-molecule polariton devices, organic light-harvesting systems, and solid-state qubits embedded in optical or microwave cavities \cite{sun2022dynamics,blais2004cavity, you2011atomic}.
	
   The main focus of this work is to investigate  the influence of phonon degrees of freedom on the  non-Markovian effects in JCH model. Non-Markovianity in open quantum systems refers to the temporary return of information from the environment to the system, resulting in partial restoration of coherence and correlations \cite{Haikka,li2018concepts,rivas2010entanglement,Breuer2016}. Such memory effects manifest as departures from monotonic decay in dynamical quantities like coherence and population imbalance \cite{Titas}. Several measures have been proposed to quantify non-Markovianity, including those based on CP-divisibility \cite{rivas2010entanglement}, trace distance \cite{BLP}, quantum Fisher information \cite{lu2010quantum,li2018concepts}, and coherence backflow via the $l_1$-norm of coherence \cite{Titas,Cramer}. In this work, we adopt a coherence-based witness, employing the $l_1$-norm of coherence to track memory effects in the qubit’s reduced dynamics under cavity and phonon couplings. The non-monotonic behavior of coherence serves as a direct, operational signature of non-Markovianity, reflecting the influence of environmental structure and system-phonon interactions \cite{douce2015quantum,dakir2025,chin2012quantum,MRB}.

	To address the strong qubit–phonon interaction analytically, we employ a unitary transformation known as Lang-Firsov transformation \cite{Lang, Tara, Mahan,Cui} that eliminates the direct qubit-phonon interaction by embedding its effects into renormalized qubit-cavity couplings in the transformed frame. We derive a time-convolutionless master equation for the qubit using the Redfield approach in the Born approximation\cite{breuer2002theory, saima,ferialdi2017exact,zhang2019exact}. The resulting dynamics feature time-dependent decay rates and energy shifts (Lamb shifts), whose explicit forms reveal how phonon-induced sidebands and cavity memory effects jointly influence the qubit coherence and population dynamics. The time evolution of the qubit density matrix and  $l_1$-norm of coherence is analyzed as a function of time \cite{Cramer}, identifying parameter regimes where coherence revivals occur, indicating non-Markovian behavior \cite{Titas,dakir2025}. 
	{It is important to distinguish the physical origin of non-Markovianity in the	present model from that in conventional spin--boson or polaronic open-system	settings. In many earlier studies, memory effects arise directly from structured	or non-Ohmic phonon environments\cite{iles2024capturing,pouthier2013reduced,gribben2022exact}. In contrast, here non-Markovianity originates from the structured photonic reservoir, whose Lorentzian spectral density provides a finite correlation time and a genuine	source of environmental memory, rather than from the phonons themselves. Phonons do not act as a memory-bearing environment but instead modify the	system’s access to cavity-induced memory through strong local dressing. This separation of roles between phonons and the cavity constitutes a qualitatively	different physical scenario and forms the central focus of this work.}
	
	The paper is organized as follows. In Section II we introduce the Jaynes–Cummings–Holstein model and describe the system–bath interactions in detail. Section III presents the dynamical evolution of the qubit in the Lang–Firsov transformed frame, where we derive the time-local master equation governing the reduced dynamics. In Section IV we discuss the non-Markovianity using the $l_1$-norm of coherence and analyze its behavior under different system and bath parameters. Finally, we conclude in section V

\section{Effective Description of JCH model}
We consider a qubit represented by a spin-1/2 particle confined to a detuned cavity  and strongly interacting with a single vibrational mode. The total Hamiltonian describing this system is known as the Jaynes–Cummings–Holstein (JCH) model \cite{shore1993jaynes,hou2024}, which combines the Jaynes–Cummings interaction for light–matter coupling \cite{jj} with the Holstein model for electron–phonon interactions \cite{holstein1959studies}. This model has been realized in various experimental platforms, including circuit QED systems \cite{wallraff2004strong}, trapped ions \cite{leibfried2003quantum}, and optomechanical setups \cite{aspelmeyer2014cavity}. The total Hamiltonian of the system is given by:
\begin{eqnarray}
	\label{OH}
	H= H_q + H_c+ H_{qc}+ H_p + H_{qp},
\end{eqnarray}
where $H_q= \frac{\omega_0}{2}\sigma_z$ is the Hamiltonian for  qubit with $\omega_0$ to be energy splitting between two levels and $\sigma_z$ is the Pauli $z$-spin matrix. The cavity Hamiltonian is $H_c= \sum_k \omega_k b_k^{\dagger}b_k$, with $\omega_k$ to be the energy of $k$th mode of the cavity and $b_k^{\dagger}, b_k$ are respectively, creation and annihilation operators for the $k$th mode. The qubit-cavity interaction is given by the Hamiltonian $H_{qc}= \sum_k( g_k \sigma^+ b_k + g_k^* \sigma^- b_k^{\dagger})$ with $g_k$ as coupling constant, $\sigma^{\pm}$ as raising (lowering) operators for the qubit. 
The cavity modes are described by the Lorentzian spectral density of the form:
\begin{eqnarray}
	J(\omega)= \frac{\gamma_0 }{2\pi}\frac{\lambda^2}{(\omega-\omega_0+\Delta)^2+ \lambda^2},
\end{eqnarray}
where $\gamma_0$ is the effective qubit-cavity coupling and $\lambda$ is the spectral width of the bath.  $\Delta=\omega-\omega_c$ is the cavity detuning with $\omega_c$ as the central frequency.
The vibrational mode of energy $\Omega$ with $ \nu^{\dagger},  \nu$ as the creation and annihilation operators respectively, is described by the Hamiltonian $H_p= \Omega \nu^{\dagger} \nu$. The qubit-phonon coupling of strength $g_p$ is given by the Hamiltonian $H_{qp}= g_p \Omega \sigma_z(\nu^{\dagger}+ \nu)$ which represents the Holstein term.

Since the qubit and phonon are strongly coupled with each other, we utilize a unitary transformation to decouple the qubit-phonon interaction $H_{qp}$ in its original frame. Such a procedure is known as Lang-Firsov (polaron) transformation\cite{Lang,Tara}. We define an operator $\mathcal{S}= -g_p \sigma_z(\nu-\nu^{\dagger})$ such that the operator $U=e^{-S}$ is unitary. Under this transformation, only qubit and phonon operators are transformed while cavity operators remain unchanged. Therefore, we write the effective Hamiltonian as
\begin{eqnarray}
	H^{LF}= U^{\dagger}HU= e^{\mathcal{S}}H e^{-\mathcal{S}}= H_q^{LF} + H_c^{LF}+ H_{qcp}^{LF}+ H_p^{LF}, 
\end{eqnarray}
where the LF transformed Hamiltonians for qubit, cavity and phonons are respectively 
$H_q^{LF}=H_q$; $ H_c^{LF}=H_c$;  $H_p^{LF}=H_p$ while the effective interaction in LF frame is given by 
\begin{eqnarray}
H_{qcp}^{LF}=e^{-2g^2_p} \sum_k( {g}_k \sigma^+ b_k \mathcal{F}+ {g}_k^* \sigma^- b_k^{\dagger}\mathcal{F}^{\dagger}),
\end{eqnarray}
with  $\mathcal{F}= e^{2g_p \nu^{\dagger}}e^{-2g_p \nu}$ is the  phonon operator in the LF frame. Therefore, effective coupling in polaron frame is reduced by a factor of $e^{-2g^2_p}$. {Although the qubit--phonon coupling can be strong in the original Hamiltonian, no perturbative approximation is applied in this bare frame. The Lang--Firsov	transformation eliminates the direct qubit--phonon interaction exactly, and its	effects are fully retained through a renormalization of the remaining	system--environment couplings. All subsequent approximations are therefore	performed only in the transformed (polaron) frame, where the effective	qubit--cavity interaction can be made weak.} 

Now we look at various energy scales that describe dynamics in the underlying model. The spectral width of cavity modes is $\lambda$ and the cavity-correlation decay scale is $\tau_c\sim \lambda^{-1}$, the relaxation time scale for the qubit is $\omega_{0}^{-1}$. Furthermore, energy scales that arise due to phonons is adiabaticity parameter given by $\frac{\gamma_0}{\Omega}$ and the qubit-phonon interaction energy scale by $\frac{g_p \Omega}{\Omega}=g_p$. The competition among all these scales govern the dynamics of the qubit. In the polaron frame, the adiabaticity parameter changes to $\frac{\gamma_0 e^{-2g^2_p}}{\Omega}$, and the condition $\frac{\gamma_0 e^{-2g^2_p}}{\Omega}<1$ is called as anti-adiabaticity.  In LF frame we perform our calculation under anti-adiabatic conditions.

\section{Dynamical evolution in LF frame}

\subsection{Initial state preparation}
We assume  initially (in original frame) an uncorrelated qubit-cavity-phonon state: $\rho_T(0)= \rho_S\otimes \rho_c\otimes \rho_p$.  $\rho_S$ as the initial state of the qubit, 
$\rho_c=\ket{0}\bra{0}$ and  $\rho_p=\ket{0}\bra{0}$ as the initial state of the cavity and phonon in vacuum state  respectively. While transforming to polaron frame, the initial state of qubit-bath can turn into an entangled state. However, under anti-adiabatic condition, we show that the total initial state can  still be represented as product state. At second order of perturbation, it is sufficient to consider the first order correction to wave function which is given 
\begin{eqnarray}
	|\varphi_m\rangle= |\varphi^0_m\rangle + \sum_{n\ne m} \frac{\langle \varphi^0_n|H_{qcp}^{LF}|\varphi^0_m\rangle}{E^0_m-E^0_n}|\varphi^0_n\rangle,
\end{eqnarray}
where $|\varphi^0_n\rangle$ represent the unperturbed (initial uncorrelated) states  with unperturbed energy $E_n^0$.  Under the condition $\frac{\gamma_0 e^{-2g^2_p}}{\Omega}<1$, we can safely ignore the second term. Therefore, upto second order of perturbation the system-bath state remains uncorrelated in LF frame.

\subsection{Dynamics from Redfield equation}
{In the polaron frame, the residual interaction between the qubit and the cavity
	is renormalized by an exponential factor $e^{-2g_p^2}$. As a consequence, the
	effective qubit--cavity coupling becomes weak for sufficiently strong phonon
	coupling. The validity of the Born approximation and the second-order
	time-local Redfield treatment is therefore controlled by the small,
	dimensionless parameter
	\begin{equation*}
		\epsilon = \frac{\gamma_0 e^{-2g_p^2}}{\Omega} \ll 1,
	\end{equation*}
	which defines the anti-adiabatic regime.
Under this condition, the reduced system--environment interaction in the
Lang--Firsov (LF) frame is weak, allowing us to employ a Redfield-type master
equation within the Born approximation and assuming an initially separable
total density operator,
$\rho_T(0) = \rho_q(0) \otimes \rho_c \otimes \rho_p$
\cite{breuer2002theory,Clos2012,Shibata1977}. We therefore obtain the
time-local non-Markovian quantum master equation in the LF frame as}
\begin{eqnarray}
	\frac{d\rho_q}{dt}=-\int_{0}^{t}d\tau{\rm Tr}_{cp}\comm{H^{LF}_{qcp}(t)}{\comm{H^{LF}_{qcp}(\tau)}{\rho_q(t)\otimes\rho_c\otimes\rho_p}}.
	\label{masmt}
\end{eqnarray}
In appendix \ref{mq}, we simplify this equation to get  amplitude damping with time dependent decay rates
\begin{eqnarray}
	\frac{d\rho_{q}}{dt}=-i\comm{\tilde{H}^{LF}_q}{\rho_{q}}+\Gamma(t)\left[2\sigma^+\rho_{q}\sigma^--\acomm{\sigma^+\sigma^-}{\rho_{q}}\right].
	\label{FE}
\end{eqnarray}
Here, $\tilde{H}^{LF}_q=H^{LF}_q-S(t)\sigma^\dagger\sigma^-$ is the renormalized Hamiltonian with $S(t)$ as the Lamb shift and $\Gamma(t)$ the time dependent decoherence function with the explicit form as given below:
\begin{eqnarray}
	\Gamma(t) &=&\frac{\gamma_0\lambda}{2}e^{-4g^2_p}\sum_{l=0}^{\infty}\frac{(4g^2_p)^l}{l!} 	\left[\frac{\lambda \left(1 - e^{-\lambda t} \cos[(\Delta + \Omega l)t]\right) + (\Delta + \Omega l)\, e^{-\lambda t} \sin[(\Delta + \Omega l)t]}{\lambda^2 + (\Delta + \Omega l)^2} \right]\\
	S(t) &=&\frac{\gamma_0\lambda}{2}e^{-4g^2_p}\sum_{l=0}^{\infty}\frac{(4g^2_p)^l}{l!}	\left[\frac{(\Delta + \Omega l) \left(1 - e^{-\lambda t} \cos[(\Delta + \Omega l)t]\right) - \lambda\, e^{-\lambda t} \sin[(\Delta + \Omega l)t]}{\lambda^2 + (\Delta + \Omega l)^2}\right]
\end{eqnarray}
{The structure of the decay rate $\Gamma(t)$ reveals explicitly how phonon
	dressing modifies information backflow in a non-Markovian photonic environment.
	Following the Lang--Firsov transformation, the decay rate contains multi-phonon
	sideband contributions, reflecting transitions at energies
	$\omega_0 + l\Omega$. These terms show that phonon dressing reshapes the
	effective spectral overlap between the qubit and the cavity across multiple
	energy scales. Importantly, the suppression of non-Markovianity emerges
	dynamically from the interplay between the cavity correlation function and the
	phonon-dressed qubit spectrum, rather than being imposed phenomenologically.
}

\begin{figure}[t]
	\includegraphics[width=5cm]{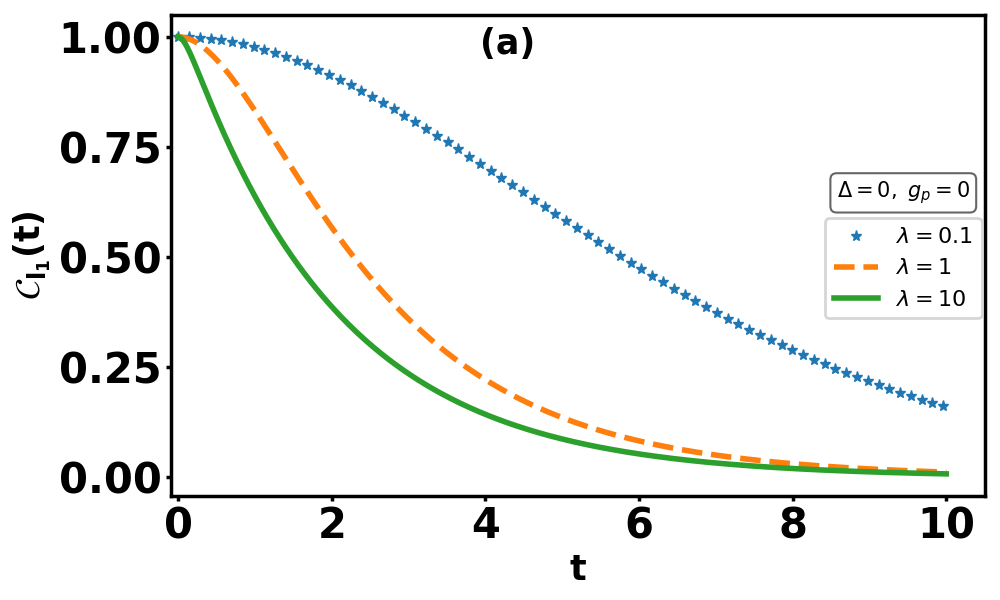}\hspace{0.1cm}
	\includegraphics[width=5cm]{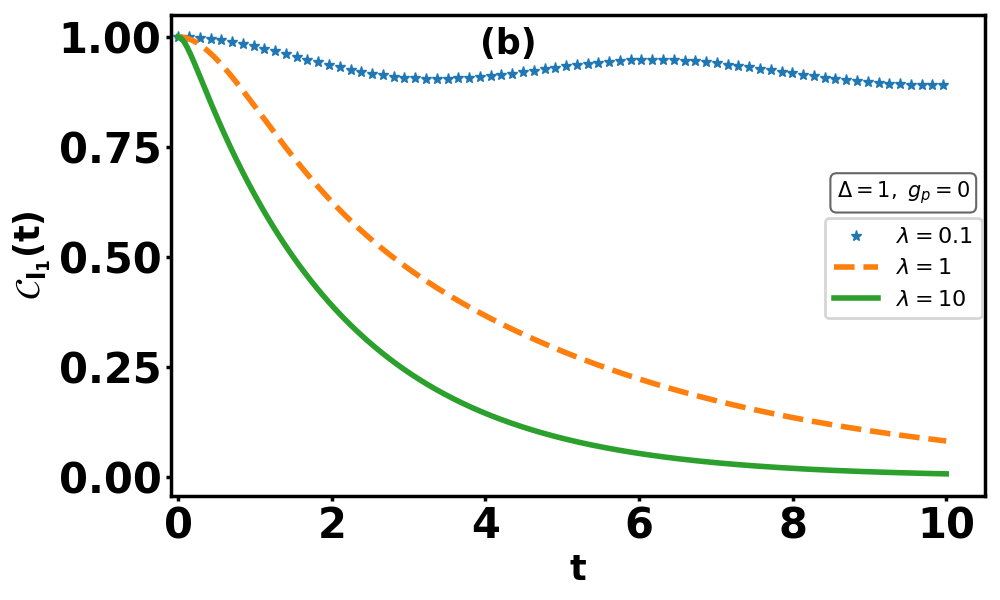}\hspace{0.1cm}
	\includegraphics[width=5cm]{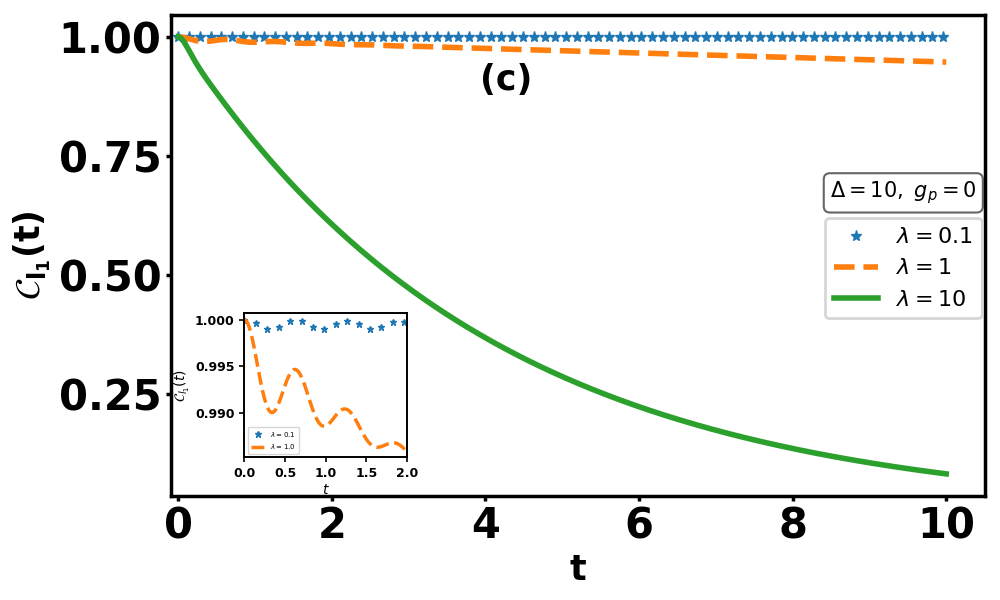}\\
	\includegraphics[width=5cm]{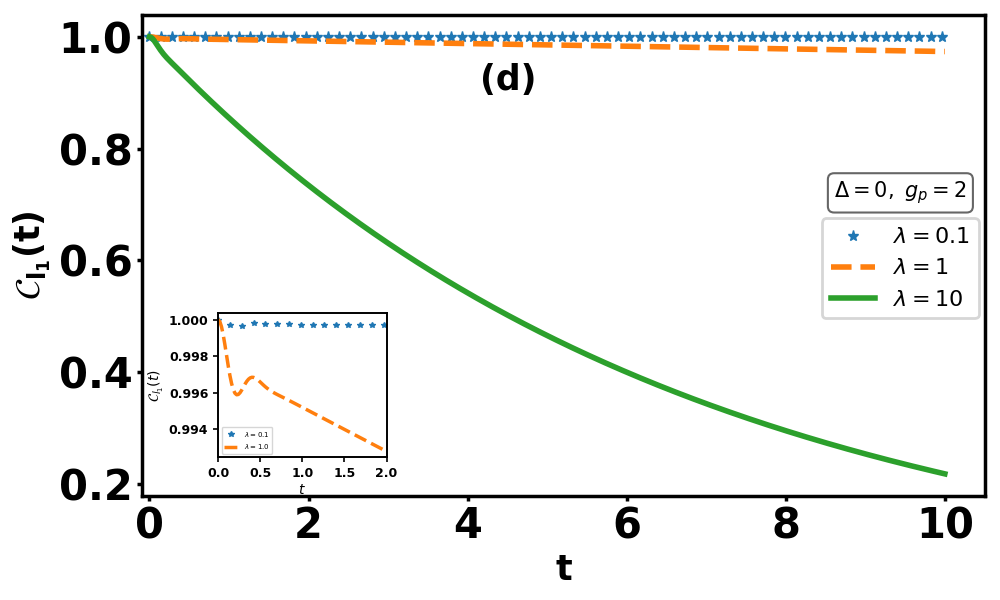}\hspace{0.1cm}
	\includegraphics[width=5cm]{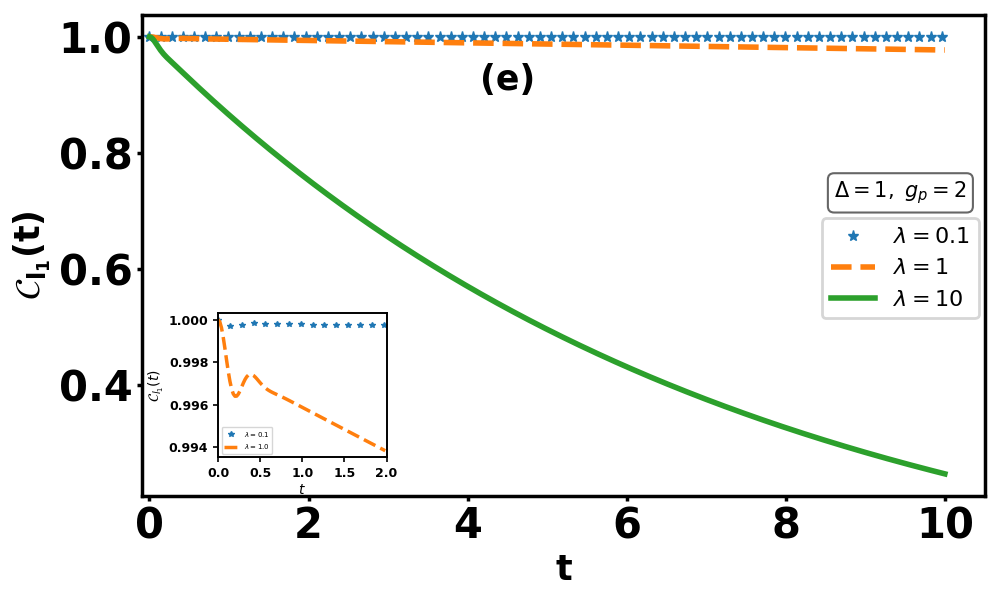}\hspace{0.1cm}
	\includegraphics[width=5cm]{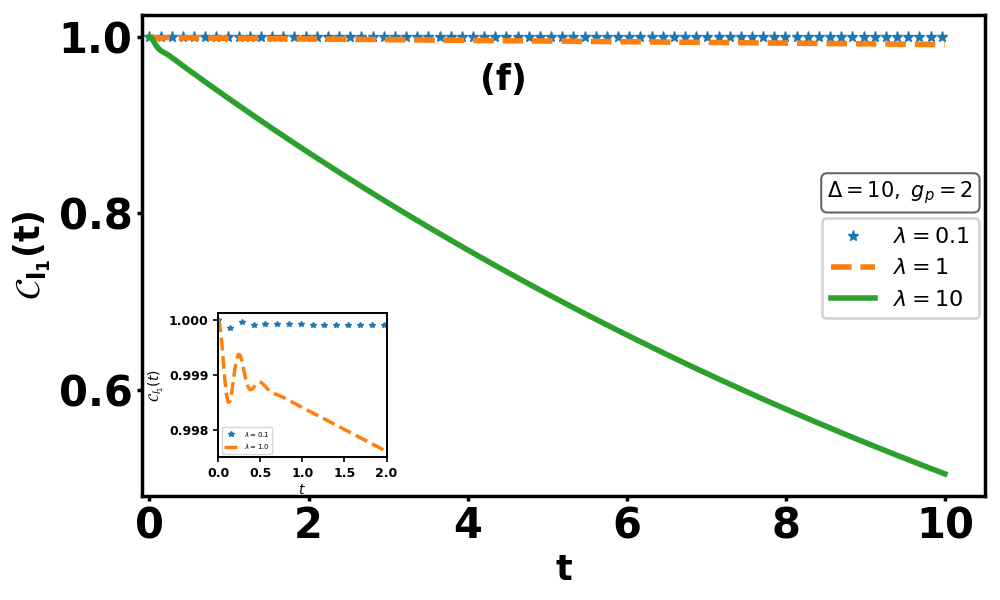}
	\caption{Time evolution of coherence $C_{l_1}(t)$ for different values of parameters. The plots from (a) to (c) represent coherence in absence of phonons i.e. $g_p=0$. The non-zero effect of phonons (i.e. $g_p=2 \ne 0$) are taken into plots from (d) to (f). The values for detuning parameter $\Delta$ are  taken as $0$, $1$ and $10$. Inset in various plots show the behaviour of coherence for short time scale. During short time evolution, coherence oscillates near the maximum value and finally saturates to maximum value in long time limit. {Notably, plots (d)-(f) show that, in the presence of phonon-induced dressing, the coherence dynamics becomes largely insensitive to detuning, indicating that phonon renormalization dominates over detuning effects in this regime.}}
	\label{coh}
\end{figure}

We now assume an initial state of the qubit to be $|\psi\rangle_q= a|0\rangle+ b|1\rangle$, then the time evolved density matrix can be calculated using above master equation and is given by
	

\begin{eqnarray}
	\rho_q(t)= \begin{pmatrix}
		|a|^2 \mathcal{G}_0(t) & a^*b\mathcal{G}_1(t)\\
		ab^*\mathcal{G}^*_1(t) & 1-|a|^2 \mathcal{G}_0(t)
	\end{pmatrix},
\end{eqnarray}
Here, $\mathcal{G}_0(t) = e^{-2\gamma(t)}$ and $\mathcal{G}_1(t) = e^{-\gamma(t) + i \Phi(t)}$, where the functions $\gamma(t)$ and $\Phi(t)$ are defined as $ \gamma(t) = \int_0^{t} ds \Gamma(s),   \Phi(t) = \int_0^{t} ds S(s)$ respectively. The dynamics corresponds to an amplitude damping channel $\rho_q(t) =\sum_i K_i \rho_q(0) K_i^{\dagger},$
	where the Kraus operators are ( $\mathcal{G}_0(t)=|\mathcal{G}_1(t)|^2$ ):
	\begin{eqnarray}
		K_0= \begin{pmatrix}
		\mathcal{G}_1(t) & 0\\
		0 & 1
	\end{pmatrix}~~~~K_1=\begin{pmatrix}
	0 & \sqrt{1-\mathcal{G}_0(t)}\\
	0& 0
		\end{pmatrix}.
	\end{eqnarray}
We now plot in  figure \ref{coh} the coherence quantified by the $l_1$-norm, $C_{l_1}(t) = \sum_{i \ne j} |\rho_{ij}(t)|=2 |a b|\, e^{-\gamma(t)}$ with respect to time $t$ for different values of parameters involved in the model. In plots from fig. \ref{coh}(a) to fig. \ref{coh}(c), we have taken $g_p=0$ i.e. bare JC model. In such case, we observe non-monotonic behaviour of coherence for narrow spectral width 
$0<\lambda <1$. This decay shows robustness for longer times as we increase detuning from $\Delta=0$ in fig. \ref{coh}(a) to $\Delta=10$ in fig. \ref{coh}(c), and finally saturated at maximum allowed value. For other values of spectral width $\lambda=10$, coherence follows monotonic decay implying Markovian dynamics. 
In fig.\ref{coh}(d) to fig.\ref{coh}(f), we incorporate phonon degrees of freedom. We observe that coherence has non-monotonic behaviour for $0<\lambda<1$ for all values of detuning. For large spectral width of the cavity modes i.e. $\lambda>1$, coherence vanishes monotonically for all values of $\Delta$ signaling Markovian dynamics.  The inset in each plot reflects the variation of coherence for small time intervals. {The qualitative similarity of	(d)–(f) plots indicates that, once phonon-induced dressing is	present, the dynamics becomes largely insensitive to detuning. Physically, the	strong phonon coupling renormalizes the effective qubit–cavity interaction,	thereby dominating the coherence evolution and suppressing detuning-induced	variations. As a result, phonon dressing governs the dynamics in this regime,	leading to robust coherence behavior in the polaron frame across a wide range of	detuning values.}

\section{Assessing Non-Markovian Dynamics}

Assessing non-Markovianity involves analyzing how a quantum system interacts with its environment and tracking the resulting evolution of its quantum states to detect departures from memoryless, Markovian behavior. The choice of a suitable non-Markovianity measure is closely tied to the nature of the system and the specifics of its coupling to the environment. Among the available techniques, information-theoretic approaches particularly those based on the dynamics of quantum coherence, have proven especially insightful. In a purely Markovian process, the coherence within a quantum system decays steadily over time, as quantum superpositions are irreversibly lost to the environment. However, non-Markovian dynamics can manifest as temporary increases or revivals of coherence, signaling a partial return of information from the environment to the system. One effective and widely used strategy for quantifying such behavior relies on monitoring the time evolution of the $l_1$-norm of coherence under incoherent, completely positive, trace-preserving (CPTP) maps, where any non-monotonic behavior serves as a clear indicator of non-Markovian effects~\cite{Titas,chen2016quantifying,xu2020detecting}. The total amount of such coherence revival, and thus the degree of non-Markovian memory effects, can be quantified by integrating the positive increments of coherence:
\begin{equation}
	\mathcal{N} = \int_{\dot{C}_{l_1}(t) > 0} \dot{C}_{l_1}(t)\, dt,
\end{equation}
where $\dot{C}_{l_1}(t)$ is the time derivative of $C_{l_1}(\rho(t))$. A vanishing $\mathcal{N}_C$ indicates purely Markovian dynamics, while a nonzero value quantifies the cumulative memory effects present in the system.


\begin{figure}[t]
	\includegraphics[width=5cm]{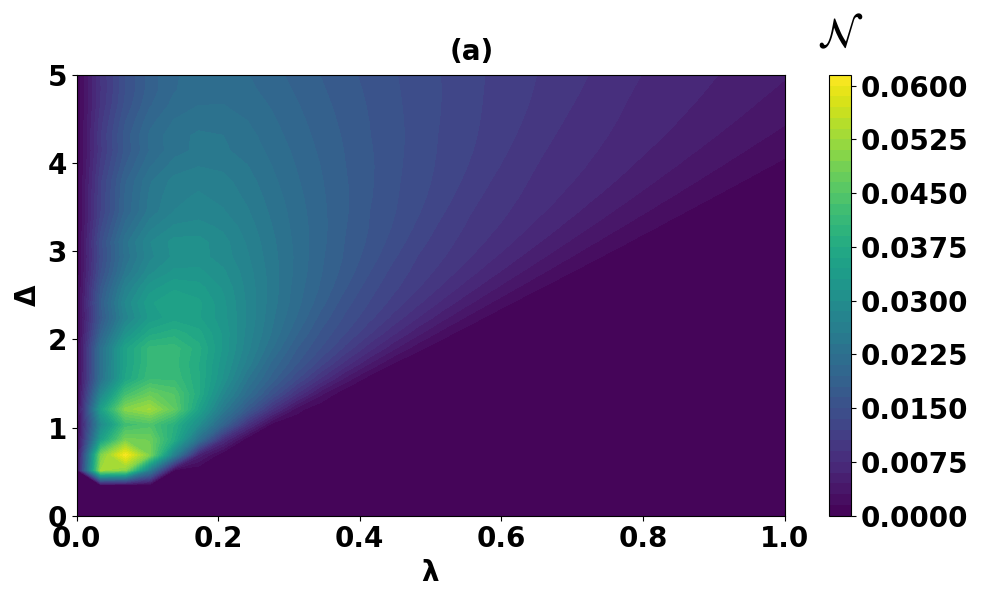} 
	\includegraphics[width=5cm]{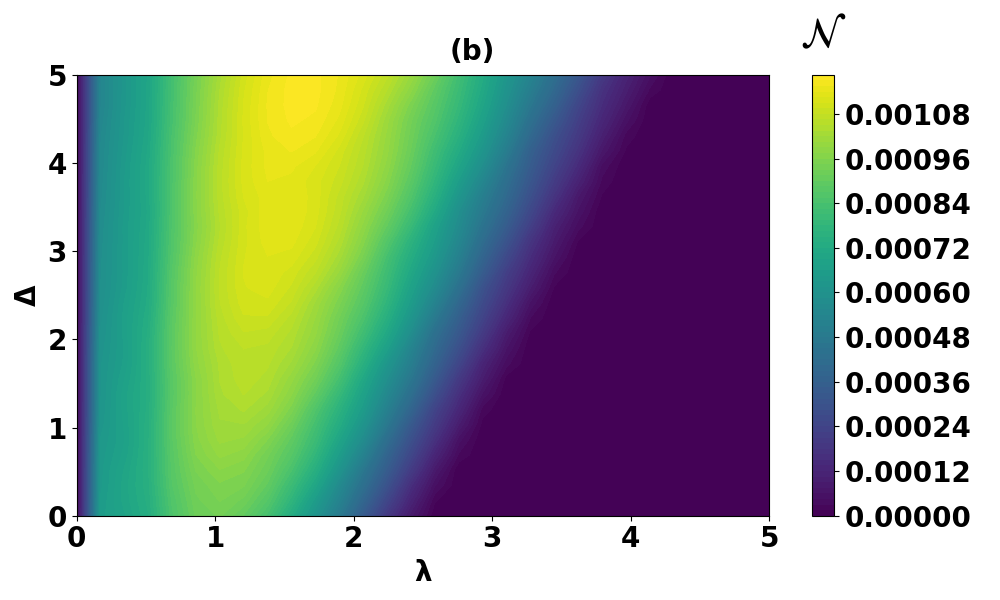} 
	\caption{
	Non-Markovianity $\mathcal{N}$ plotted as a function of detuning $\Delta$ and bath spectral width $\lambda$. Figure (a) corresponds to the case without phonon coupling ($g_p=0$), where pronounced non-Markovian behavior is observed across a range of parameters due to memory effects from the structured photonic environment. In contrast, Figure (b) shows that when phonon coupling is introduced ($g_p \neq 0$), non-Markovian features are significantly suppressed, as the phonon-induced dressing effectively inhibits coherence backflow and reduces memory effects.}
		\label{figNM1}
\end{figure}
\begin{figure}[ht]
	\includegraphics[width=5cm]{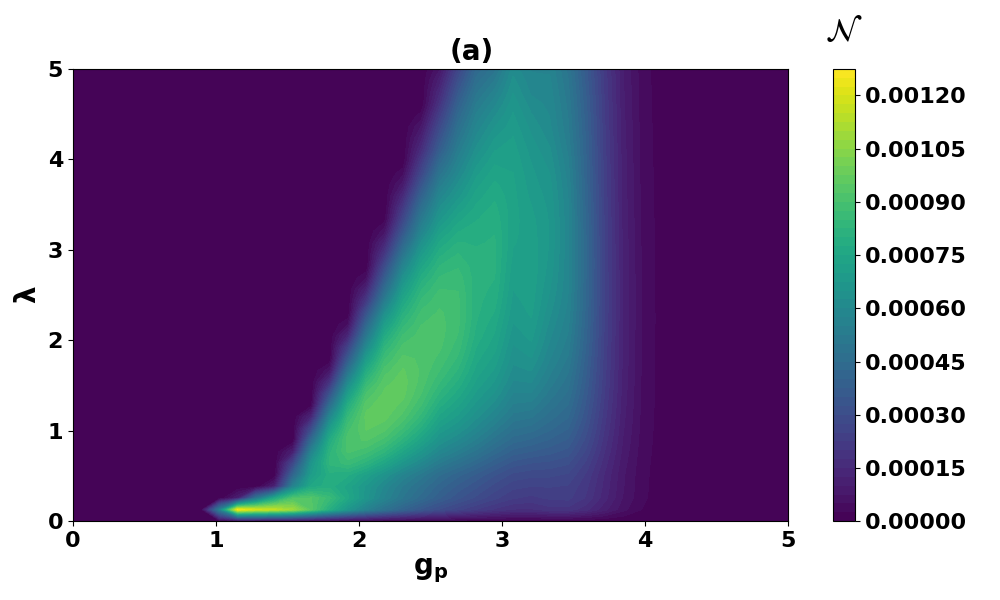}\hspace{0.1cm}
	\includegraphics[width=5cm]{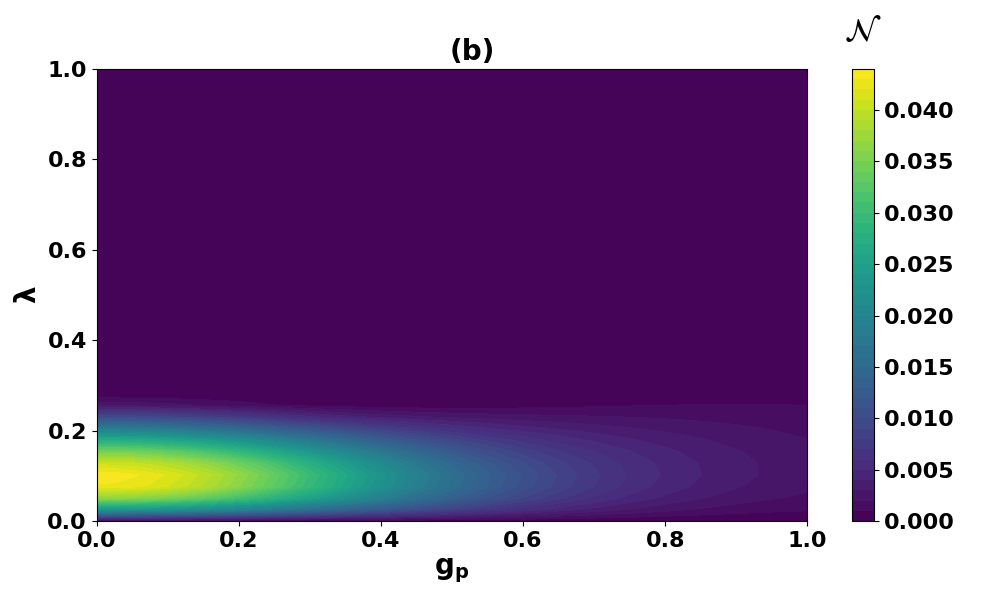}\hspace{0.1cm}
	\includegraphics[width=5cm]{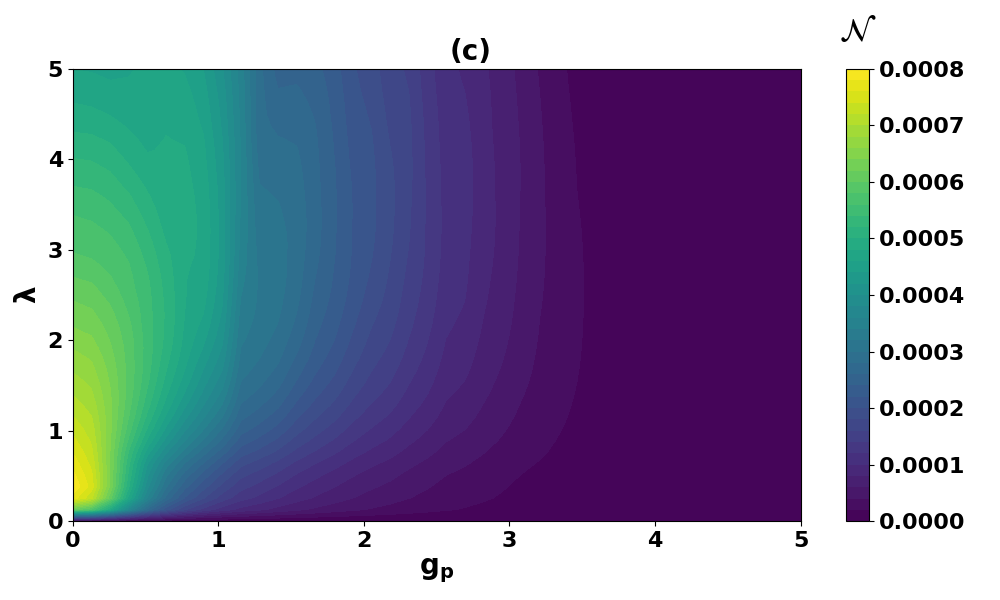}
\caption{Non-Markovianity $\mathcal{N}$ as a function of the bath spectral width $\lambda$ and the qubit–phonon coupling strength $g_p$ for different detuning ($\Delta$) values. Figure (a) corresponds to $\Delta = 0$, where $\mathcal{N}$ remains finite within a narrow range of $g_p$. Figure (b) represents moderate detuning ($\Delta = 1$), where non-Markovianity becomes localized for smaller values of $\lambda$ and $g_p$, while figure (c) shows the large-detuning case ($\Delta = 10$), where $\mathcal{N}$ increases with $\lambda$ for weak phonon coupling ($g_p < 1$).}
	\label{figNM2}
\end{figure}

In figure \ref{figNM1}(a)-(b), we plot non-Markovianity $\mathcal{N}$ as a function of $\lambda$ and $\Delta$ for the model under consideration. The non-Markovianity in case of bare JC model (see fig.\ref{figNM1}(a)) is localized within a narrow range of spectral width $0<\lambda<1$. This is attributed to different processes that occur at second order of perturbation in polaron frame. In the bare Jaynes–Cummings (JC) model (Fig.~\ref{figNM1}a), $\mathcal{N}$ is significant only for narrow cavity spectral widths ($0 < \lambda < 1$), since a narrow spectrum corresponds to a long cavity correlation time ($\tau_c \sim \lambda^{-1}$), enabling the qubit to partially recover information previously lost to the cavity. Broader spectra lead to faster decay of cavity correlations, suppressing non-Markovian effects. Incorporating strong qubit-phonon interactions in the Jaynes–Cummings–Holstein (JCH) model (Fig.~\ref{figNM1}(b)) reduces the peak values of $\mathcal{N}$ because phonon dressing, via the Lang-Firsov transformation, renormalizes the qubit-cavity coupling by $e^{-2 g_p^2}$, limiting energy exchange and coherence backflow. Simultaneously, phonon dressing redistributes the qubit spectral weight, allowing memory effects to persist across a wider range of spectral widths $\lambda$ and qubit-phonon couplings $g_p$. {The observed order-of-magnitude suppression of non-Markovianity at strong phonon coupling reflects the emergence of a distinct dynamical regime rather than a trivial rescaling of known behavior (compare  Fig.~\ref{figNM1}(a) and \ref{figNM1}(b)). In the bare Jaynes--Cummings limit, long	cavity correlation times directly translate into pronounced coherence revivals	and strong information backflow. In contrast, under strong phonon dressing, the	cavity correlations remain long-lived, but the dressed qubit couples to them only	weakly due to the exponential renormalization of the effective qubit--cavity	interaction. As a result, environmental memory persists in principle but becomes	dynamically inaccessible to the system.}  
{ Importantly, phonon dressing plays a dual role in the non-Markovian dynamics. While it strongly suppresses the magnitude of non-Markovianity, it simultaneously allows cavity-induced memory effects to survive over a broader range of cavity spectral widths. In standard single-bath polaronic models, increasing coupling typically drives the system monotonically toward Markovian behavior. Here, however, the non-Markovian window in parameter space persists even under strong phonon dressing, albeit with substantially reduced amplitude. This nontrivial redistribution of memory effects arises from the coexistence of two environments with competing roles: phonons act as local dressing agents that renormalize system couplings, whereas the cavity serves as the primary source of environmental memory.}

Next, we plot in figure \ref{figNM2}(a)-(c), $\mathcal{N}$ with respect to $\lambda$ and $g_p$ for different values of $\Delta$. 
We observe that detuning $\Delta$ further modulates these trends: at zero detuning ($\Delta=0$), resonance enhances qubit-cavity exchange, so non-Markovianity is finite but confined to a narrow range of $g_p$; at moderate detuning ($\Delta = 1$), memory effects require both long cavity correlation (small $\lambda$) and weak phonon dressing (small $g_p$), localizing $\mathcal{N}$ to a restricted parameter space; at large detuning ($\Delta = 10$), non-Markovianity increases with $\lambda$ for weak $g_p (< 1)$, as broadening the cavity spectrum partially restores spectral overlap with the qubit and allows backflow of information. {Thus the role of detuning further confirms that the suppression of non-Markovianity is	not a mere quantitative artifact. At strong phonon coupling, the qubit spectrum	is redistributed into phonon sidebands at energies $\omega_0 + l\Omega$. Finite	detuning allows the cavity spectral density to overlap selectively with these	sidebands, reopening indirect, phonon-assisted exchange channels. This mechanism	accounts for the partial restoration of non-Markovianity observed at large	detuning, even when the bare qubit--cavity coupling is strongly suppressed.} These observations reflect the competition between cavity correlation time, effective qubit-cavity coupling renormalized by phonons, and detuning-induced resonance, with the polaron-frame second-order perturbative analysis capturing the subtle effects of phonon dressing and anti-adiabatic conditions on qubit dynamics.
\section{Conclusion}
{
	In conclusion, we have analyzed the polaron effects on the non-Markovian dynamics of a qubit in a cavity in presence of strong local vibrational mode given by JCH model. We employ LF transformation to eliminate the qubit-phonon interaction in the transformed frame while its influence emerges as renormalization of qubit-cavity couplings. 
	This procedure enables an analytical treatment of the reduced qubit dynamics in the anti-adiabatic regime using a time-local Redfield master equation with explicitly time-dependent decay rates and Lamb shifts.
	
	The results obtained demonstrate that the non-Markovian effects in this hybrid system is controlled by cavity correlations that enable the information backflow while phonon-induced dressing  controls the qubit's effective access to this  memory. In the absence of phonons, narrow cavity spectra give rise to long-lived reservoir correlations and pronounced coherence oscillations and revivals, which can be further enhanced by detuning. When strong qubit–phonon coupling is present, the formation of a local phonon cloud exponentially suppresses the effective qubit–cavity interaction, leading to a substantial reduction of coherence revivals and non-Markovianity. Importantly, this suppression reflects a dynamical inhibition of memory access rather than a loss of environmental correlations. For broad cavity spectra, rapid decay of reservoir correlations results in monotonic coherence decay and effectively Markovian dynamics, irrespective of phonon coupling.
	
	Furthermore, detuning plays a nontrivial role by redistributing the qubit spectral weight into phonon sidebands, enabling phonon-assisted spectral overlap with the cavity modes. This mechanism can partially restore non-Markovian features even when the bare qubit–cavity coupling is strongly renormalized, highlighting the competition between cavity correlation time, phonon-induced renormalization, and detuning-controlled spectral overlap in governing information backflow.

	The parameters of the present model admit a natural correspondence with experimentally tunable quantities across several platforms. In circuit QED, the cavity spectral width is determined by the resonator linewidth, detuning is controlled via qubit biasing, and effective phonon-like couplings arise from substrate modes or engineered electromechanical elements \cite{blais2021circuit,chu2017quantum}. In organic and molecular cavity-QED systems, strong vibronic coupling is intrinsic, with phonon frequencies and coupling strengths fixed by molecular structure, while cavity losses and detuning govern the photonic environment \cite{wang2024dispersive,herrera2017dark}. Trapped-ion and optomechanical platforms offer a particularly clean realization, where vibrational couplings and structured reservoirs can be tuned with high precision \cite{barreiro2011open,muller2012engineered}. Across these settings, the present results indicate that while the photonic environment determines the presence of environmental memory, phonon-induced dressing governs the system’s ability to exploit that memory.
	}

\section{Acknowledgemnets}
SB acknowledges the hospitality at QD Lab, Department of Physics, University of Kashmir, where most of this work was done. SB also acknowledges financial support from DST, Government of India, under the INSPIRE Fellowship no. IF210401. 

\section{Declarations}
\textbf{Data Availability}: No datasets were generated or analysed in the current study.

\textbf{Competing Interests}: The authors declare no competing interests.

\appendix
\section{Effective Hamiltonian}
In this appendix, we provide the derivation of effective Hamiltonian in LF frame. 
We start with the original Hamiltonian in equation \ref{OH}: 
\begin{eqnarray}
	\label{H}
	H=\omega_{0}\sigma^{z}+\sum_{k} b^\dagger_kb_k+\sum_{k}(g_k\sigma^\dagger b_k+g^\star \sigma^- b^\dagger_k)+\Omega\nu^\dagger \nu+g_p\Omega\sigma_{z}(\nu+\nu^\dagger)
\end{eqnarray}
We now apply Lang-Firsov transformation via $U=e^{-S}$ with $S=-g\sigma_{z}(\nu-\nu^\dagger)$, such that the transformed Hamiltonian can be written as:
\begin{eqnarray}
	H^{LF}=U^\dagger H U=e^{S} H e^{-S}
	\label{LF}
\end{eqnarray}
Under this transformation, we see that cavity operators do not change and $\sigma_z$ also remains unchanged, that is 
 $s^{S}b_k e^{-S}=b_k$; $e^{S}\sigma_{z} e^{-S}=\sigma_{z}$. Next, we write 
\begin{eqnarray}
	e^{S}\sigma^+ e^{-S}&=&\sigma^+ +\comm{S}{\sigma^+}+\frac{1}{2}\comm{S}{\comm{S}{\sigma^+}}+...\nonumber\\
	&=&\sigma^+-g_p(\nu-\nu^\dagger)\comm{\sigma_{z}}{\sigma^+}+...\nonumber\\
	&=&\sigma^+ -g_p(\nu-\nu^\dagger)2\sigma^+ +...\nonumber\\
\implies	e^{S}\sigma^{+} e^{-S}&=&\sigma^+ \exp[-2g_p(\nu-\nu^\dagger)].
	\label{sig}
\end{eqnarray}
Also, $e^S \nu e^{-S}$ can be written as
\begin{eqnarray}
	e^S \nu e^{-S}&=&\nu+\comm{S}{\nu}+...\nonumber\\
	&=&\nu+[-g_p\sigma_{z}(\nu-\nu^\dagger),\nu]\nonumber\\
\implies	e^S \nu e^{-S}&=&\nu-g_p\sigma_{z}.
	\label{nu}
\end{eqnarray}
Using equations \ref{sig} and \ref{nu},  we write
\begin{eqnarray}
	e^S\left[\Omega\nu^\dagger\nu+g_p\Omega\sigma_{z}(\nu+\nu^\dagger)\right]e^{-S}
	&=&\Omega[\nu^\dagger-g_p\sigma_{z}][\nu-g_p\sigma_{z}]+g_p\Omega\sigma_{z}[\nu-g_p\sigma_{z}+\nu^\dagger-g_p\sigma_{z}]\nonumber\\
	&=&\Omega\nu^\dagger\nu+g^2_p\Omega\sigma^2_{z}-g_p\Omega\sigma_{z}(\nu+\nu^\dagger)+g_p\Omega\sigma_{z}(\nu+\nu^\dagger)-2g_p^2\Omega\sigma^2_{z}\nonumber\\
	&=&\Omega\nu^\dagger\nu-g^2_p\Omega, \\
	e^S\sum_{k}\left(g_k\sigma^+ b_k+g^\star\sigma^- b^\dagger_k\right)e^{-S}\nonumber
	&=&\sum_{k}g_ks^S\sigma^+ e^{-S}b_k+h.c\nonumber\\
	&=&\sum_{k}\left(g_k \sigma^+ e^{2g_p(\nu-\nu^\dagger)}b_k+h.c\right)\nonumber\\
	&=&e^{-2g^2_p}\sum_{k}\left(g_k\sigma^+  e^{2g_p\nu^\dagger}e^{-2g_p\nu}b_k+h.c\right)\nonumber\\
	&=&e^{-2g^2_p}\sum_{k}\left(g_k\sigma^+ b_k \mathcal{F}+g^\star\sigma^- b^\dagger_k \mathcal{F^\dagger}\right),
\end{eqnarray}
where $h.c.$ means Hermitian conjugate and  $\mathcal{F}=e^{2g_p\nu^\dagger}e^{-2g_p\nu}$ represents effective phonon operator in LF frame and we have used $e^{A+B}=e^A e^B e^{\frac{-1}{2}\comm{A}{B}}$
for $\comm{A}{B}={\rm constant}$. Clubbing all the equations together, we write effective Hamiltonian in LF frame as
\begin{eqnarray}
	H^{LF}=H^{LF}_q+H^{LF}_c+H^{LF}_p+H^{LF}_{qcp}
\end{eqnarray}
where $H^{LF}_q=\frac{\omega_{0}}{2}\sigma_{z}$, $H^{LF}_c=\sum_{k}\omega_k b^\dagger_k b_k$ and $ H^{LF}_p=\Omega\nu^\dagger \nu$. The effective interaction in the LF frame is
\begin{eqnarray}
	H^{LF}_{qcp}=e^{-2g^2_p}\sum_{k}\left(g_k\sigma^+ b_k \mathcal{F}+g^\star\sigma^- b^\dagger_k \mathcal{F^\dagger}\right)
\end{eqnarray}

\section{Master Equation}\label{mq}
In this appendix, we provide the derivation of master equation of the main text. We start with time convolutionless master equation in the LF frame:
\begin{eqnarray}
	\frac{d\rho_q}{dt}=-\int_{0}^{t}d\tau{\rm Tr}_{c,p}\comm{H^{LF}_{qcp}(t)}{\comm{H^{LF}_{qcp}(\tau)}{\rho_q(t)\otimes\rho_c\otimes\rho_p}}.
	\label{meq}
\end{eqnarray}
 Here, we  have assumed factorized initial state of the total system:  $\rho_T(0)=\rho_{q}\otimes\rho_{c}\otimes\rho_{p}$ with $\rho_{q}, \rho_{c} ,\rho_{p}$ to be the density operators for the qubit,cavity and phonons respectively and  $\rm Tr_{c,p}$ is the trace  taken over cavity and phonon modes.  $H^{LF}_{qcp}(t)$ is defined in interaction picture as follows:
\begin{eqnarray}
	\label{11}
	H^{LF}_{qcp}(t)&=&e^{i(H^{LF}_q+H^{LF}_c+H^{LF}_p)t} H^{LF}_{qcp} e^{-i(H^{LF}_q+H^{LF}_c+H^{LF}_p)t}\nonumber\\
	&=&e^{-2 g^2_{p}}\sum_{k}\left[g_{k}\sigma^{+} b_{k} e^{i(\omega_{0}-\omega_{k}) t}\mathcal{F}(t)+g^{\star}_{k}\sigma^{-} b^\dagger_{k} e^{i(\omega_{0}-\omega_{k})t}\mathcal{F}^{\dagger}(t)\right],
\end{eqnarray}
where $ \mathcal{F}(t)= e^{2g_p\nu^\dagger e^{i\Omega t}}e^{-2g_p\nu e^{-i\Omega t}} $. Next, we evaluate the double commutator in the above master equation \ref{meq}. The first term gives:
\begin{eqnarray}
	{\rm Tr}_{cp}  \left[H^{LF}_{qcp}(t)H^{LF}_{qcp}(\tau)\rho_{q}(t)\rho_{c}\rho_{p}\right]&=& e^{-4g^2_p} \sum_{k}\sum_{k^\prime} \Bigg[ g_k g^{\star}_{k^\prime} \sigma^{+} \sigma^{-} \rho_{q}(t) e^{i(\omega_{0}-\omega_{k})t} e^{-i(\omega_{0}-\omega_{k^\prime})\tau} \langle b_k b^{\dagger}_{k^\prime}\rangle_c \langle\mathcal{F}(t)\mathcal{F}^{\dagger}(\tau)\rangle_p \nonumber \\
	&&~~~~~~~~~~~~~~ +~ g^\star_k g_{k^\prime} \sigma^{-} \sigma^{+} \rho_{q}(t) e^{-i(\omega_{0}-\omega_{k})t} e^{i(\omega_{0}-\omega_{k^\prime})\tau} \langle b^\dagger_k b_{k^\prime}\rangle_c \langle\mathcal{F}^\dagger(t)\mathcal{F}(\tau)\rangle_p \Bigg].
\end{eqnarray}
Here, we have used $\sigma^{+2}=0=\sigma^{-2}$. Also, we assume state of the cavity as $\rho_c=|0\rangle\langle0|$ with $|0\rangle$ to be the vacuum state. Therefore, 
$\langle b^\dagger_k b_{k^\prime}\rangle_c=0$ and $\langle b_k b^{\dagger}_{k^\prime}\rangle_c=\delta_{k k^\prime}$ and above equation simplifies to 
\begin{eqnarray}
	\label{ist}
		{\rm  Tr}_{cp}\left[H^{LF}_{qcp}(t)H^{LF}_{qcp}(\tau)\rho_{q}(t)\rho_{c}\rho_{p}\right]
	&=&e^{-4g^2_p}\sum_{k}\left[|g_k|^2 e^{i(\omega_{0}-\omega_{k})(t-\tau)}\langle\mathcal{F}(t)\mathcal{F}^{\dagger}(\tau)\rangle_p\sigma^+\sigma^{-}\rho_{q}(t)\right]. 
\end{eqnarray}
Similarly, the second term of equation \ref{meq} can be written as
\begin{eqnarray}
	\label{2nd}
{\rm Tr}_{cp}\left[H^{LF}_{qcp}(t)\rho_{q}(t)\rho_{c}\rho_{p}H^{LF}_{qcp}(\tau)\right]
	&=&e^{-4g^2_p}\sum_{k}\left[|g_k|^2 e^{-i(\omega_{0}-\omega_{k})(t-\tau)}\langle\mathcal{F}(\tau)\mathcal{F}^{\dagger}(t)\rangle_p\sigma^{+}\rho_{q}(t)\sigma^{-}\right].
\end{eqnarray}
 
We define the total correlation function of cavity and phonon modes as
\begin{eqnarray}
C(t-\tau)=e^{-4g^2_p}\sum_{k}\left[|g_k|^2 e^{-i(\omega_{0}-\omega_{k})(t-\tau)}\langle\mathcal{F}(\tau)\mathcal{F}^{\dagger}(t)\rangle_p\right]
\end{eqnarray}
Collecting all these terms and transforming back to Schrodinger picture, we get the master equation 
\begin{eqnarray}
	\frac{d\rho_{q}}{dt}=-i\comm{\tilde{H}^{LF}_q}{\rho_{q}}+\Gamma(t)\left[2\sigma^{+}\rho_{q}\sigma^{-}-\acomm{\sigma^+\sigma^-}{\rho_{q}}\right]
\end{eqnarray}
Where $\tilde{H}^{LF}_q=H^{LF}_q-\int_{0}^{t}d\tau \imaginary[ C(t-\tau)]~\sigma^+ \sigma^-$ is the renormalized Hamiltonian and the time dependent decoherence function is given by
\begin{eqnarray}
\Gamma(t)=\int_{0}^{t}\real [C(t-\tau)]=e^{-4g^2_p}\sum_{k}\int_{0}^{t}d\tau \real\left[|g_k|^2 e^{-i(\omega_{0}-\omega_{k})(t-\tau)}\langle\mathcal{F}(\tau)\mathcal{F}^{\dagger}(t)\rangle_p\right]
\end{eqnarray}

Next we can calculate correlation functions
\begin{eqnarray}
	\langle\mathcal{F}(t)\mathcal{F}^{\dagger}(\tau)\rangle_p&=&\bra{0} e^{2g_p\nu^\dagger e^{i\Omega t}}e^{-2g_p\nu e^{-i\Omega t}} e^{-2g_p\nu^\dagger e^{i\Omega \tau}}e^{2g_p\nu e^{-i\Omega \tau}}\ket{0}\nonumber\\
	&=&\bra{0} e^{-2g_p\nu e^{-i\Omega t}}e^{-2g_p\nu^\dagger e^{i\Omega \tau}}\ket{0}\nonumber\\
	&=&\bra{0}e^{-2g_p\nu^\dagger e^{i\Omega \tau}}e^{2g_p\nu^\dagger e^{i\Omega \tau}} e^{-2g_p\nu e^{-i\Omega t}}e^{-2g_p\nu^\dagger e^{i\Omega \tau}}\ket{0}\nonumber\\
	&=&\bra{0}e^{-2g_pe^{-i\Omega t}(\nu-2g_p e^{i\Omega \tau})}\ket{0}\nonumber\\
	&=&e^{4g^2_p e^{-i\Omega(t-\tau)}}
\end{eqnarray}
Similarly, its conjugate is $	\langle\mathcal{F}(\tau)\mathcal{F}^{\dagger}(t)\rangle_p=e^{4g^2_p e^{i\Omega(t-\tau)}}$.
Since the cavity mode are described by lorentz spectral density
\begin{eqnarray}
	J(\omega)=\frac{\gamma_0\lambda^2}{2\pi}\frac{1}{(\omega-\omega_{0})^2+\lambda^2}
\end{eqnarray}
such that 
\begin{eqnarray}
	\sum_{k}|g_k|^2 e^{-i(\omega_{0}-\omega_{k})(t-\tau)}&=&\int d\omega J(\omega)e^{-i(\omega_{0}-\omega)(t-\tau)}\nonumber\\
	&=&\frac{1}{2}\gamma_0\lambda e^{-(\lambda-i\Delta)(t-\tau)}
\end{eqnarray}
Therefore, we write
\begin{eqnarray}
	\Gamma(t)&=&\real\left[e^{-4g^2_p}\sum_{k}\int_{0}^{t}d\tau|g_k|^2 e^{-i(\omega_{0}-\omega_{k})(t-\tau)}\langle\mathcal{F}(\tau)\mathcal{F}^{\dagger}(t)\rangle_p\right]\nonumber\\
	&=&\real\left[\frac{\gamma_0\lambda}{2}e^{-4g^2_p}\int_{0}^{t}d\tau e^{-(\lambda-i\Delta)(t-\tau)}e^{4g^2_p e^{i\Omega(t-\tau)}}\right]\nonumber\\
	&=&\real\left[\frac{\gamma_0\lambda}{2}e^{-4g^2_p}\sum_{l=0}^{\infty}\frac{(4g^2_p)^l}{l!}\int_{0}^{t}d\tau e^{-(\lambda-i\Delta)(t-\tau)}e^{i\Omega l(t-\tau)}\right]\nonumber\\ 
	&=&\real\left[\frac{\gamma_0\lambda}{2}e^{-4g^2_p}\sum_{l=0}^{\infty}\frac{(4g^2_p)^l}{l!} \frac{1 - e^{-\left[\lambda - i(\Delta +\Omega l)\right] t}}{\lambda - i(\Delta + \Omega l)}\right] 
\end{eqnarray}
\begin{eqnarray}
		\Gamma(t) &=&\frac{\gamma_0\lambda}{2}e^{-4g^2_p}\sum_{l=0}^{\infty}\frac{(4g^2_p)^l}{l!} 	\left[\frac{\lambda \left(1 - e^{-\lambda t} \cos[(\Delta + \Omega l)t]\right) + (\Delta + \Omega l)\, e^{-\lambda t} \sin[(\Delta + \Omega l)t]}{\lambda^2 + (\Delta + \Omega l)^2} \right]\\
	S(t) &=&\frac{\gamma_0\lambda}{2}e^{-4g^2_p}\sum_{l=0}^{\infty}\frac{(4g^2_p)^l}{l!}	\left[\frac{(\Delta + \Omega l) \left(1 - e^{-\lambda t} \cos[(\Delta + \Omega l)t]\right) - \lambda\, e^{-\lambda t} \sin[(\Delta + \Omega l)t]}{\lambda^2 + (\Delta + \Omega l)^2}\right]
\end{eqnarray}

\section{Relation between coherence and trace-distance non-Markovianity}

{In this appendix, we analyze the relationship between the coherence-based non-Markovianity measure adopted in the main text and the trace-distance criterion for the dynamics generated by the Jaynes–Cummings–Holstein model. The reduced qubit dynamics in Eq. \ref{FE} takes the form of a time-local amplitude-damping channel characterized by a decoherence function

\begin{eqnarray}
	\mathcal{G}_1(t) = e^{-\gamma(t) + i\Phi(t)}, \qquad 
	\gamma(t) = \int_0^t ds\, \Gamma(s).
\end{eqnarray}
\textit{\textbf{{Coherence dynamics:}}}
The off-diagonal element of the reduced density matrix evolves as
\begin{eqnarray}
	\rho_{01}(t) = \rho_{01}(0)\, \mathcal{G}_1(t),
\end{eqnarray}
so that the $l_1$-norm of coherence is
\begin{eqnarray}
	C_{l_1}(t) = 2|\rho_{01}(0)|\, e^{-\gamma(t)}.
\end{eqnarray}
Therefore, a temporary increase of coherence,
$\dot C_{l_1}(t) > 0$, occurs if and only if $\Gamma(t) < 0$.

\textit{\textbf{Trace-distance dynamics:}} The trace distance between two qubit states,
$D(t)=\frac{1}{2}\|\rho_1(t)-\rho_2(t)\|$,
for amplitude-damping channels is maximized by an optimal pair of states lying
on the equatorial plane of the Bloch sphere. For such states, the trace distance
evolves as
\begin{eqnarray}
	D(t) = D(0)\, |\mathcal{G}_1(t)| = D(0)\, e^{-\gamma(t)}.
\end{eqnarray}
Hence, trace-distance backflow ($\dot D(t)>0$) also occurs if and only if
$\Gamma(t)<0$.

Thus, for the amplitude-damping dynamics induced by the Jaynes–Cummings–Holstein model, both coherence backflow and trace-distance backflow originate from the same physical mechanism, namely the temporary negativity of the time-dependent decay rate $\Gamma(t)$. Within this class of dynamics, coherence revivals and trace-distance increases constitute equivalent indicators of non-Markovian behavior.}

\bibliography{jaynes.bib}
\end{document}